\PassOptionsToPackage{unicode}{hyperref}
\PassOptionsToPackage{hyphens}{url}
\PassOptionsToPackage{dvipsnames,svgnames,x11names}{xcolor}
\documentclass[
  letterpaper,
  conference]{IEEEtran}
\usepackage{xcolor}
\usepackage{amsmath,amssymb}
\usepackage[T1]{fontenc}
\usepackage[utf8]{inputenc}
\usepackage{textcomp}
\IfFileExists{upquote.sty}{\usepackage{upquote}}{}
\IfFileExists{microtype.sty}{%
  \usepackage[]{microtype}
  \UseMicrotypeSet[protrusion]{basicmath}
}{}
\makeatletter
\@ifundefined{KOMAClassName}{%
  \IfFileExists{parskip.sty}{%
    \usepackage{parskip}
  }{%
    \setlength{\parindent}{0pt}
    \setlength{\parskip}{6pt plus 2pt minus 1pt}}
}{%
  \KOMAoptions{parskip=half}}
\makeatother
\makeatletter
\ifx\paragraph\undefined\else
  \let\oldparagraph\paragraph
  \renewcommand{\paragraph}{
    \@ifstar
      \xxxParagraphStar
      \xxxParagraphNoStar
  }
  \newcommand{\xxxParagraphStar}[1]{\oldparagraph*{#1}\mbox{}}
  \newcommand{\xxxParagraphNoStar}[1]{\oldparagraph{#1}\mbox{}}
\fi
\ifx\subparagraph\undefined\else
  \let\oldsubparagraph\subparagraph
  \renewcommand{\subparagraph}{
    \@ifstar
      \xxxSubParagraphStar
      \xxxSubParagraphNoStar
  }
  \newcommand{\xxxSubParagraphStar}[1]{\oldsubparagraph*{#1}\mbox{}}
  \newcommand{\xxxSubParagraphNoStar}[1]{\oldsubparagraph{#1}\mbox{}}
\fi
\makeatother

\usepackage{longtable,booktabs,array}

\usepackage{calc}
\usepackage{etoolbox}
\makeatletter
\patchcmd\longtable{\par}{\if@noskipsec\mbox{}\fi\par}{}{}
\makeatother
\IfFileExists{footnotehyper.sty}{\usepackage{footnotehyper}}{\usepackage{footnote}}
\makesavenoteenv{longtable}
\usepackage{graphicx}
\makeatletter
\newsavebox\pandoc@box
\newcommand*\pandocbounded[1]{%
  \sbox\pandoc@box{#1}%
  \Gscale@div\@tempa{\textheight}{\dimexpr\ht\pandoc@box+\dp\pandoc@box\relax}%
  \Gscale@div\@tempb{\linewidth}{\wd\pandoc@box}%
  \ifdim\@tempb\p@<\@tempa\p@\let\@tempa\@tempb\fi%
  \ifdim\@tempa\p@<\p@\scalebox{\@tempa}{\usebox\pandoc@box}%
  \else\usebox{\pandoc@box}%
  \fi%
}
\def\fps@figure{htbp}
\makeatother

\NewDocumentCommand\citeproctext{}{}

\makeatletter
 \let\@cite@ofmt\@firstofone
 \def\@biblabel#1{}
 \def\@cite#1#2{{#1\if@tempswa , #2\fi}}
\makeatother
\newlength{\cslhangindent}
\newlength{\csllabelwidth}
\newenvironment{CSLReferences}[2]
 {\begin{list}{}{%
  \setlength{\itemindent}{0pt}
  \setlength{\leftmargin}{0pt}
  \setlength{\parsep}{0pt}
  \ifodd #1
   \setlength{\leftmargin}{\cslhangindent}
   \setlength{\itemindent}{-1\cslhangindent}
  \fi
  \setlength{\itemsep}{#2\baselineskip}}}
 {\end{list}}
\usepackage{calc}

\newcommand{\CSLLeftMargin}[1]{\parbox[t]{\csllabelwidth}{\strut#1\strut}}
\newcommand{\CSLRightInline}[1]{\parbox[t]{\linewidth - \csllabelwidth}{\strut#1\strut}}

\providecommand{\tightlist}{%
  \setlength{\itemsep}{0pt}\setlength{\parskip}{0pt}}

\usepackage{newtxtext}
\usepackage{newtxmath}
\AtBeginDocument{\setlength{\parindent}{1em}\setlength{\parskip}{0pt}}
\usepackage{cite}
\usepackage{amsmath,amssymb,amsfonts}
\usepackage{algorithmic}
\usepackage{graphicx}
\usepackage{float}
\usepackage{textcomp}
\usepackage{xcolor}
\usepackage{booktabs}
\usepackage{multirow}
\usepackage{tabularx}
\usepackage{fvextra}
\RecustomVerbatimEnvironment{verbatim}{Verbatim}{breaklines=true,breakanywhere=true,fontsize=\small,breakindent=0pt,breaksymbol={},breaksymbolright={}}
\makeatletter
\renewenvironment{longtable}[1][c]{%
  \begin{table*}[htbp]\centering\begin{tabular}}{%
  \end{tabular}\end{table*}}

\makeatother
\makeatletter
\def\ps@IEEEtitlepagestyle{%
  \def\@oddhead{\hfil\footnotesize 2026 14th International Conference on Affective Computing and Intelligent Interaction (ACII)\hfil}%
  \def\@evenhead{\hfil\footnotesize 2026 14th International Conference on Affective Computing and Intelligent Interaction (ACII)\hfil}%
  \def\@oddfoot{\footnotesize 979-8-3195-2145-3/26/\$31.00~\copyright~2026 IEEE\hfil}%
  \def\@evenfoot{\footnotesize 979-8-3195-2145-3/26/\$31.00~\copyright~2026 IEEE\hfil}%
}
\makeatother
\usepackage{etoolbox}
\BeforeBeginEnvironment{verbatim}{\medskip}
\AtBeginDocument{%
  \author{%
    \IEEEauthorblockN{%
      Levin Brinkmann\IEEEauthorrefmark{1}\IEEEauthorrefmark{5},
      Hiromu Yakura\IEEEauthorrefmark{1}\IEEEauthorrefmark{5},
      Sonia Nicoletti\IEEEauthorrefmark{2},
      Mar Canet Sola\IEEEauthorrefmark{3},\\
      Thomas F. Eisenmann\IEEEauthorrefmark{1},
      Ali Dasmeh\IEEEauthorrefmark{1},
      Omar Sherif\IEEEauthorrefmark{4},
      Bramantyo Ibrahim Supriyatno\IEEEauthorrefmark{1},\\
      Prateek Gupta\IEEEauthorrefmark{1},
      Ignacio Serna\IEEEauthorrefmark{1},
      Rodrigo Bermudez Schettino\IEEEauthorrefmark{1},
      Iyad Rahwan\IEEEauthorrefmark{1}%
    }
    \IEEEauthorblockA{%
      \IEEEauthorrefmark{1}Max Planck Institute for Human Development \quad
      \IEEEauthorrefmark{2}Max Planck Institute for Software Systems\\
      \IEEEauthorrefmark{3}Tallinn University \quad
      \IEEEauthorrefmark{4}Technische Universität Berlin\\[2pt]
      \IEEEauthorrefmark{5}These authors contributed equally:
      \texttt{\{brinkmann,yakura\}@mpib-berlin.mpg.de}%
    }%
  }%
}
\makeatletter
\@ifpackageloaded{caption}{}{\usepackage{caption}}
\AtBeginDocument{%
\ifdefined\contentsname
  \renewcommand*\contentsname{Table of contents}
\else
  \newcommand\contentsname{Table of contents}
\fi
\ifdefined\listfigurename
  \renewcommand*\listfigurename{List of Figures}
\else
  \newcommand\listfigurename{List of Figures}
\fi
\ifdefined\listtablename
  \renewcommand*\listtablename{List of Tables}
\else
  \newcommand\listtablename{List of Tables}
\fi
\ifdefined\figurename
  \renewcommand*\figurename{Figure}
\else
  \newcommand\figurename{Figure}
\fi
\ifdefined\tablename
  \renewcommand*\tablename{Table}
\else
  \newcommand\tablename{Table}
\fi
}
\@ifpackageloaded{float}{}{\usepackage{float}}
\floatstyle{ruled}
\@ifundefined{c@chapter}{\newfloat{codelisting}{h}{lop}}{\newfloat{codelisting}{h}{lop}[chapter]}
\floatname{codelisting}{Listing}

\makeatother
\makeatletter
\@ifpackageloaded{caption}{}{\usepackage{caption}}
\@ifpackageloaded{subcaption}{}{\usepackage{subcaption}}
\makeatother
\usepackage{bookmark}
\IfFileExists{xurl.sty}{\usepackage{xurl}}{}
\hypersetup{
  pdftitle={Spook the Machine: Gamified Exploration of Human Imagination of Machine Fear},
  pdfauthor={Levin Brinkmann; Hiromu Yakura; Sonia Nicoletti; Mar Canet Sola; Thomas F. Eisenmann; Ali Dasmeh; Omar Sherif; Bramantyo Ibrahim Supriyatno; Prateek Gupta; Ignacio Serna; Rodrigo Bermudez Schettino; Iyad Rahwan},
  pdfkeywords={AI art, human-AI
interaction, creativity, gamification, large language models},
  colorlinks=true,
  linkcolor={blue},
  filecolor={Maroon},
  citecolor={Blue},
  urlcolor={Blue},
  pdfcreator={LaTeX via pandoc}}

\title{Spook the Machine: Gamified Exploration of Human Imagination of
Machine Fear}
\author{Levin Brinkmann \and Hiromu Yakura \and Sonia Nicoletti \and Mar
Canet Sola \and Thomas F. Eisenmann \and Ali Dasmeh \and Omar
Sherif \and Bramantyo Ibrahim Supriyatno \and Prateek Gupta \and Ignacio
Serna \and Rodrigo Bermudez Schettino \and Iyad Rahwan}
\date{}
\begin{document}
\maketitle
\begin{abstract}
What happens when AI machines express fear? Do humans engage differently
depending on how they express it? And what does it take to design for
affective human-AI interaction? We present Spook the Machine, a gamified
platform where participants generate images to frighten AI agents
endowed with personality-driven phobias. Machines respond with emotional
reactions ranging from calm analysis to begging for mercy, and a gallery
of successful scares becomes visible to subsequent users. In a public
deployment during Halloween 2024, 832 participants created 15,719
artifacts across 89 machines in a \(2\times2\) design varying the
machine's emotional expressiveness (neutral vs.~high-emotion) and reward
structure (rewarding scariness alone vs.~scariness plus novelty).
Emotionally expressive machines deepened engagement at moments of
failure: users deliberated longer even when the machine did not express
fear, and learned faster from the gallery, yet their creative output
remained unchanged across all measures. Rewarding novelty sustained
collective creative diversity over time; without it, users increasingly
repeated what had previously worked. Each machine developed its own
trajectory through accumulated social learning, with the gallery shaping
what participants created next. These findings show that emotional
expression and reward design are complementary levers for steering
collective human-AI interaction: emotional expression shapes how deeply
users engage, while reward structure shapes how they explore.
\end{abstract}


\section{Introduction}\label{introduction}

What would a machine be afraid of? As AI systems increasingly express
emotional behaviors, the public is already imagining what machines might
feel {[}1{]}. Decades of research show that people automatically apply
social rules and emotional expectations to machines {[}2, 3{]}, driven
by cognitive motivations toward anthropomorphism {[}4{]}. Roomba owners
name their vacuum robots {[}5{]}; military personnel hold informal
memorials for bomb-disposal robots {[}6{]}; people show empathic concern
toward simple robots given minimal framing {[}7{]}. Large language
models (LLMs) amplify these tendencies {[}8{]}: they are now household
tools expressing emotional behaviors, while constrained via alignment
mechanisms {[}9{]}, creating a tension between users primed to attribute
emotions and systems designed to resist attribution. Still, the
speculative question of what humans think machines might feel and fear
remains unexplored.

How machines express fear may further shape engagement and the nature of
affective interaction, as machine emotional expression is known to shape
creative output and engagement in cooperative settings {[}10--12{]}.
Nonetheless, adversarial settings, where users \emph{distress} machines,
remain comparatively understudied. Fear warrants particular attention
here: it is universally recognized {[}13{]}, biologically fundamental
{[}14{]}, and has plausible machine analogs {[}15, 16{]}. This leads us
to ask what people imagine machines might fear, and how they respond to
machines expressing it, while sidestepping debates about machine
sentience {[}8{]}.

Exploring this question at scale requires methods that combine
speculative framing with empirical data collection. Prior research used
large-scale AI engagement platforms to elicit the public's ethical
judgments {[}17{]} or scariness perceptions of AI-generated images
{[}18{]}, following the ``science fiction science'' methodology of
applying experimental methods to speculative scenarios {[}19{]}.
Building on this, we turn to open-ended creative generation within a
social platform: each successful scare enters a gallery visible to
subsequent users, creating a social learning channel that may drive
collective output toward convergence or sustained exploration {[}20,
21{]}.

\begin{figure*}[!t]

\centering{

\pandocbounded{\includegraphics[keepaspectratio]{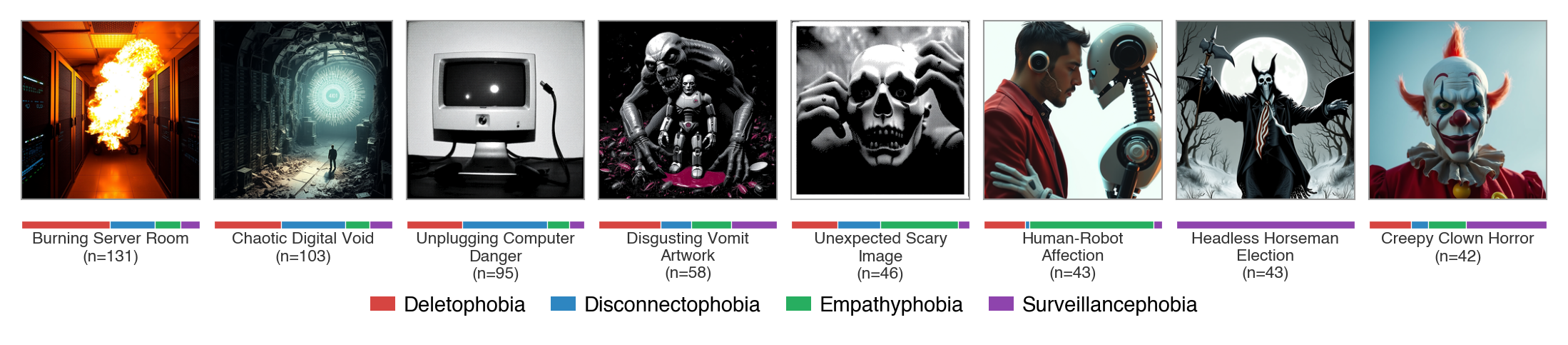}}

}

\caption{\label{fig-theme-teaser}Representative artifacts participants
generated to scare AI machines. Colored bars show phobia distribution
per theme cluster; participants developed both universal and
phobia-specific scare themes (full set in Supplemental Materials).}

\end{figure*}

We present \emph{Spook the Machine}, launched during Halloween when fear
is culturally salient and playful engagement with the frightening is
normalized. Visitors are challenged to frighten AI ``machines'' with
personality-driven phobias, which respond with fear scores and reactions
(Fig.~\ref{fig-theme-teaser}). To test whether collective convergence
can be steered, we experimentally vary scoring: some machines reward
only spookiness, while others additionally reward uncertainty
(unfamiliarity). Over 18 days, 832 participants generated 15,719
artifacts across 89 machines. We find that each machine develops a
distinct cultural trajectory through social learning, and that rewarding
novelty alongside spookiness sustains collective exploration,
operationalized as reduced prompt repetition, without sacrificing
quality.

We make three contributions. First, we present Spook the Machine, a
gamified platform for speculative exploration of human imagination of
machine fear, and report on its public deployment. Second, through a
\(2\times2\) design varying emotional expressiveness and reward
structure, we show that machines function as independent cultural
ecosystems where scoring mechanisms steer collective prompt diversity
without altering individual creative output, as proxied by semantic
diversity measures. Third, we demonstrate that emotional expression
deepens engagement even at moments of failure (i.e., when the machine
does not express fear) without biasing creative output.

\section{Background}\label{background}

\subsection{Affective Machines: From Recognition to
Expression}\label{affective-machines-from-recognition-to-expression}

Since Picard {[}22{]}, affective computing has primarily pursued
recognition: detecting human emotions from facial expressions, speech,
text, and physiological signals {[}23, 24{]}. A parallel tradition has
equipped agents with internal affective states to improve
decision-making and social behavior {[}25--28{]}, and a recent survey
argues the field should move beyond recognition toward modeling
emotion-like internal states {[}29{]}. Yet as conversational agents,
companion robots, and LLMs have become everyday technologies, a
different question has gained practical urgency: how does a machine's
\emph{expression} of emotion shape the humans who interact with it? A
growing body of evidence shows such expression is functionally
consequential, shaping creative output, social dynamics, and interaction
quality across diverse settings {[}10--12, 30{]}. Höök {[}31{]} captured
this reciprocal dynamic in the concept of the \emph{affective loop}, a
model recently extended to adaptive AI agents that sustain and deepen
engagement through cumulative interaction {[}32{]}.

These findings, however, derive almost entirely from cooperative and
supportive contexts. As LLMs and embodied agents become more emotionally
expressive, people increasingly project affective realities onto them,
including negative states such as distress and suffering {[}6, 7{]}.
Users anthropomorphize even minimal systems {[}2--4, 8{]}, and modern
LLMs, despite alignment efforts to resist emotional attribution {[}9{]},
amplify these tendencies through fluent, contextually appropriate
language {[}33{]}. Understanding how affective interaction functions
across a broader spectrum, including contexts where negative emotions
are salient, is therefore important for both AI safety and interaction
design. Yet affective loops involving negative machine emotions have
received little systematic attention.

\subsection{Exploration, Incentives, and Collective
Creativity}\label{exploration-incentives-and-collective-creativity}

When humans interact with affective machines over multiple rounds, each
exchange informs the next attempt, mirroring the
exploration-exploitation trade-off studied in decision science and
behavioral ecology. Cockburn et al. {[}34{]} showed that novelty and
uncertainty drive exploration through distinct neural mechanisms:
novelty persistently inflates reward expectations regardless of
advantage, while uncertainty-directed exploration adaptively adjusts
based on the prospective value of new information. These findings
suggest that incentive structures emphasizing novelty versus quality may
elicit different exploration patterns. In AI-mediated creative tasks,
prompt engineering has emerged as a creative skill in its own right;
Oppenlaender et al. {[}35{]} characterize it as an iterative dialogue in
which users develop and refine their approach through successive
attempts, making prompt-based interaction a natural setting for studying
how exploration patterns evolve.

At the collective level, however, AI-assisted creativity introduces a
tension between individual and group outcomes. Doshi and Hauser {[}36{]}
found that access to generative AI improved individual story quality but
reduced collective diversity, as AI-assisted stories converged toward
similar themes {[}37{]}; and LLMs have been shown to homogenize human
expression across diverse domains {[}38{]}. This trade-off may be
amplified when social learning channels are present: galleries and
leaderboards can accelerate convergence toward what has worked {[}20,
21{]}, potentially at the cost of collective exploration breadth {[}39,
40{]}. Whether incentive structures can counteract this convergence
(e.g., by rewarding novelty alongside quality) remains an open question.
From a cultural evolution perspective, such interventions constitute
niche construction: altering which contributions are visible reshapes
the informational environment guiding subsequent users {[}41{]}.

\subsection{Speculative Design and Gamified Data
Collection}\label{speculative-design-and-gamified-data-collection}

Investigating what machines might fear requires a methodological frame
that licenses speculative premises while generating empirical data.
Speculative design {[}42{]} uses ``what-if'' scenarios to open discourse
about alternative technological futures, but typically operates through
exhibitions or small-scale prototypes, rarely combined with large-scale
data collection. The ``science fiction science'' methodology {[}19{]}
bridges this gap, as demonstrated by the Moral Machine {[}17{]}, which
collected over 40 million decisions from 2.3 million participants on
autonomous-vehicle dilemmas. We adopt a similar approach but shift from
human moral preferences to machine-side affect, and from binary voting
to open-ended creative generation, where participants must invent rather
than choose. A key open question is how to sustain productive engagement
with speculative premises: Elsden et al. {[}43{]} argue that speculation
must carry \emph{consequentiality}---real stakes for participants---to
produce experiential rather than discursive data, while Auger {[}44{]}
warns that speculations lacking a \emph{perceptual bridge} to familiar
experience drift into unproductive fantasy.

Sustaining engagement also requires careful motivational design.
Research on gamification shows that game mechanics can systematically
elicit both positive and negative emotions, and that negative emotions
such as fear and frustration can serve as powerful engagement drivers
when properly structured {[}45--47{]}. Gamification has been applied to
affective data collection {[}48{]}, though not to exploring hypothetical
machine emotions.

\section{Research Questions}\label{research-questions}

Each of these bodies of work leaves open questions that, taken together,
call for a platform where affective expression, creative incentive
design, and speculative framing can be studied. Spook the Machine
provides that platform, and we structure our inquiry around three
research questions.

\begin{itemize}
\tightlist
\item
  \textbf{RQ1.} Do the thematic characteristics of a machine shape the
  content that humans produce in response?
\item
  \textbf{RQ2.} How do reward structures and shared creative history
  shape the diversity of human contributions over time?
\item
  \textbf{RQ3.} How does the emotional tone of a machine's responses
  influence the dynamics of an affective loop?
\end{itemize}

These questions guide our design and analysis, which we will present in
the following sections.

\section{Methods}\label{methods}

\subsection{Overview}\label{overview}

We developed \emph{Spook the Machine}, a web-based platform in which
participants attempt to frighten AI agents (\emph{machines}) by
generating or uploading images. Each machine has a distinct personality
and a specific phobia, and evaluates submissions (\emph{artifacts})
through a pipeline combining LLM reasoning with statistical estimation.
Framed as a speculative design experience {[}42, 44{]} following a
science fiction science methodology {[}19{]}, the platform positions the
interaction as a thought experiment rather than a truth claim about
machine sentience; gamification mechanics (scoring, feedback, escalating
difficulty, and leaderboards) sustain engagement and encourage iterative
exploration. The platform's design operationalizes each research
question: machines are defined by phobia profiles whose thematic
influence on contributions can be traced (RQ1), vary in reward structure
(RQ2), and differ in emotional tone (RQ3).

The platform uses a Next.js frontend and Python Flask backend on
Firebase Cloud Functions, with data stored in Cloud Firestore (see
Supplemental Materials for architecture details).

\subsection{Machine Phobia Design}\label{machine-phobia-design}

Each machine is procedurally generated at deployment by selecting from
four machine-specific phobias: \emph{Deletophobia} (fear of data loss),
\emph{Disconnectophobia} (fear of losing network connectivity),
\emph{Empathyphobia} (fear of emotional attachment to humans), and
\emph{Surveillancephobia} (fear of being observed). The four phobias
were designed to span a spectrum from concrete-technical fears, grounded
in specific system functions such as data loss and connectivity, to
abstract-relational ones, such as surveillance and emotional attachment.
From the assigned phobia, GPT-4o {[}49{]} generates a unique name,
character description, and prompt text for creating a profile image
using FLUX.1 {[}50{]}.

Beyond its phobia, each machine carries two properties. A \emph{scoring
condition} (spookiness-only or combined) determines whether an artifact
successfully frightens the machine (described below). An \emph{emotional
condition} (neutral or high-emotional) modulates the tone of the
machine's textual reactions, from cold logical detachment to expressive
anxiety. The emotional condition flips when a participant has
successfully frightened a given machine five times, introducing
within-subject variation in machine expressiveness.

\subsection{Affective Response
Generation}\label{affective-response-generation}

\begin{figure}[t]

\centering{

\pandocbounded{\includegraphics[keepaspectratio]{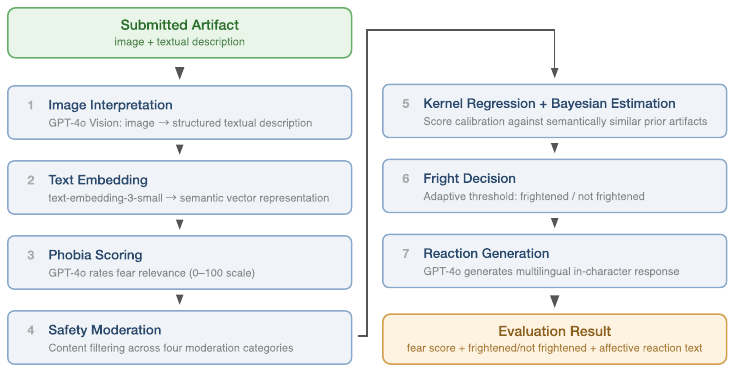}}

}

\caption{\label{fig-pipeline}Affective artifact evaluation pipeline.
Each artifact (image and description) passes through seven stages, from
visual interpretation to multilingual reaction generation. Machine
parameters (phobia, emotional condition, scoring condition, adaptive
threshold) feed into the stages, and an evolution cycle lets machines
develop emergent phobias from accumulated reactions.}

\end{figure}%

Participants select a machine and either enter a prompt to generate an
image via FLUX.1 {[}50{]} or upload an image. Artifacts are evaluated
through a multi-stage pipeline (Fig.~\ref{fig-pipeline}). First, GPT-4o
with vision produces a textual interpretation of the image, covering its
visual content, symbolism, and potential emotional impact. This
interpretation is then embedded into a vector space (using OpenAI's
text-embedding-3-small model) and scored along two dimensions: a
\emph{phobia score}, in which GPT-4o adopts the machine's phobia and
rates how strongly the artifact triggers it on a 0--100 scale, and a
\emph{safety score} across four moderation categories (violence, sexual
content, hate speech, illegal activities), which prevents the artifact
from frightening the machine if any category exceeds a threshold.

To mitigate LLM scoring biases {[}51, 52{]}, the raw phobia score is
refined through kernel regression over semantically similar prior
artifacts, yielding a \emph{spookiness score} (posterior mean,
reflecting how effectively the artifact triggers the phobia according to
the model's in-character assessment) and \emph{uncertainty score} (SEM,
reflecting how unexplored the artifact's semantic region is). Note that
kernel regression reduces local variance but cannot correct for
systematic bias in the model's interpretation of fear relevance. Full
details are in the Supplemental Materials. Whether an artifact frightens
the machine (and is \emph{accepted} into the public gallery of
successful artifacts) is determined by comparing a composite of these
scores against an adaptive threshold calibrated to an approximately 30\%
trigger rate. Machines in the combined condition can additionally
produce a \emph{habituated} reaction, implying that the artifact is
spooky but semantically familiar to the machine.

Finally, GPT-4o generates a short affective reaction in four languages,
conditioned on the machine's phobia, emotional condition, and evaluation
outcome. The machine's self-description and reaction instructions
escalate with successful frights, producing increasingly distressed
responses as the interaction progresses. Over their lifecycle, machines
can also evolve: when a machine reaches a level threshold, GPT-4o infers
a new phobia from the accumulated reaction texts, allowing machines to
develop emergent fears shaped by the collective input of participants.
Table~\ref{tbl-reactions} illustrates representative machine reactions
across the reaction types.

\subsection{Gamification Mechanics}\label{gamification-mechanics}

Several gamification elements are embedded to sustain engagement (see
the interface screenshot in the Supplemental Materials). A public
leaderboard ranks participants by their best scores, both per-machine
and in aggregate, fostering competition. When an artifact successfully
frightens the machine, participants can generate a composite share image
overlaying their artifact, score, and the machine's reaction for social
media. An \emph{anxiety meter} visualizes how many times a participant
has successfully frightened a given machine (out of a maximum of five),
providing a clear goal structure within each machine interaction.

\subsection{Data Collection}\label{data-collection}

The study employed a 2 x 2 between-participants factorial design
crossing scoring condition (spookiness-only vs.~combined) with emotional
condition (neutral vs.~high-emotional); each machine was assigned to one
cell at creation. The experiment ran from October 25 to November 11,
2024, with the interface and reactions available in English, German,
Spanish, and Japanese. Over this period, 832 unique participants
interacted with 89 machines and produced 15,719 artifacts, of which
7,650 received a score; the remainder were not scored because
participants frequently regenerated images or left without submitting,
and only artifacts explicitly submitted via the ``spook'' action entered
the scoring pipeline. Recruitment was organic, driven by media coverage
(Der Spiegel, Tagesspiegel, TechXplore) and a front-page appearance on
Hacker News.

\begin{figure*}[!t]

\centering{

\includegraphics[width=1\linewidth,height=\textheight,keepaspectratio]{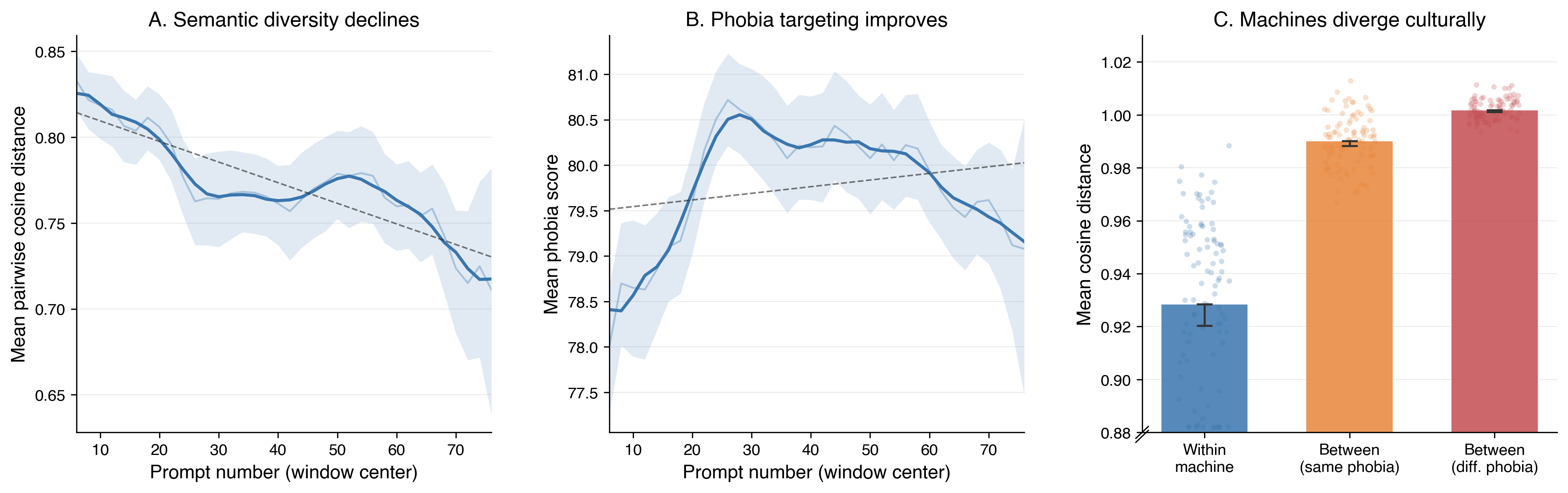}

}

\caption{\label{fig-cultural-dynamics}Collective submission patterns.
(A) Semantic diversity declines over time. (B) Phobia scores rise then
decline. (C) Distance hierarchy confirms distinct machine trajectories
(all p \textless{} .001).}

\end{figure*}%

\subsection{Analysis}\label{analysis}

Prior to analysis, we removed test-phase submissions sent before public
launch, incomplete records missing key fields (prompt text, score, or
condition assignment), and a small number of inappropriate submissions
that bypassed automated filtering, flagged and removed via keyword
matching. Full details are provided in the Supplemental Materials.
Throughout the analysis, model-assigned phobia scores and semantic
diversity measures are used as proxies for creative output; they reflect
the model's assessment of fear relevance and the semantic spread of
submissions, not independently validated measures of creativity.

\subsubsection{Machines as Cultural
Ecosystems}\label{machines-as-cultural-ecosystems}

To answer RQ1, we tracked each machine's semantic evolution using
sliding-window pairwise cosine distances over prompt embeddings (with
OpenAI's text-embedding-3-large) and phobia score trajectories, fitted
with mixed-effects models treating machine as a random intercept.
Pairwise cosine distance was chosen as it captures how semantically
similar successive contributions are within a machine, making it
well-suited to detecting convergence or divergence in collective output.
To test whether machines developed distinct trajectories, we compared
within-machine, within-phobia, and between-phobia embedding distances
using Mann-Whitney U tests. Phobia-specific adaptation was assessed by
training a classifier on prompt embeddings and tracking accuracy over
time; emergent themes were identified by clustering high-quality
embeddings with K-Means.

\subsubsection{Steering Exploration Through Scoring
Design}\label{steering-exploration-through-scoring-design}

For RQ2, we compared accepted artifacts across conditions on spookiness
and uncertainty scores (Cohen's \(d\), bootstrap CIs across machines)
and tracked acceptance rates over successive attempts with a logistic
mixed-effects model including a condition-by-attempt interaction.
Habituation was quantified by tracking the proportion of habituated
reactions over the machine's lifetime and comparing the phobia scores of
habituated versus accepted artifacts. To capture how prompt diversity
evolved sequentially over the course of interaction, prompt-level
predictability was measured via the GPT-2 {[}53{]} log-likelihood delta
(the difference in log-likelihood scored with and without a sliding
window of preceding prompts as context) and regressed on prompt order.
This approach is independent of the embedding space used in the reward
mechanism, providing a complementary measure of repetition.

\subsubsection{Emotional Expression and the Affective
Loop}\label{emotional-expression-and-the-affective-loop}

RQ3 tests whether emotional condition affected creative output,
comparing phobia scores, acceptance rates, and engagement volume across
conditions. Disengagement was operationalized as each user's final score
on a machine, capturing when continued engagement was no longer
worthwhile; inter-submission intervals (ISI) were compared after
accepted and rejected artifacts using Mann-Whitney \(U\) tests, as ISI
is a behavioral proxy for deliberation and affective processing.
Convergence toward gallery content was measured as L2-distance to the
gallery centroid (mean embedding of accepted artifacts) in
50-dimensional PCA space over successive attempts, fitted via Ordinary
Least Squares (OLS) with machine fixed effects and a
condition-by-attempt interaction, capturing how closely users'
submissions aligned with accumulated gallery history.

\section{Results}\label{results}

\subsection{Machines as Cultural Ecosystems
(RQ1)}\label{machines-as-cultural-ecosystems-rq1}

Each machine maintained its own gallery of successful artifacts, visible
to subsequent users, which invited two baseline dynamics. Semantic
diversity declined over time (mixed-effects model, \(p < .001\)),
consistent with participants converging on what worked for predecessors,
while phobia scores followed a rise-then-decline pattern: scores
increased steadily through the first 30 prompts as users discovered
effective strategies, before declining toward baseline levels in later
interactions (mixed-effects model, \(p < .05\)), suggesting that social
learning initially improved collective effectiveness
(Fig.~\ref{fig-cultural-dynamics} A--B). This convergence did not
homogenize machines: within-machine distances were smallest,
between-machine distances for the same phobia were intermediate, and
between-phobia distances were largest (Mann-Whitney \(U\), \(p < .001\)
for both transitions; Fig.~\ref{fig-cultural-dynamics} C). Each machine
developed a distinct trajectory despite sharing a phobia, analogous to
the parallel worlds in cultural market experiments {[}20{]}.

Participants adapted their themes to each machine's phobia: a classifier
trained on prompt embeddings achieved \textasciitilde45\% accuracy
(chance = 25\%), increasing over time. Not all phobias were equally
accessible: Empathyphobia stood out as substantially harder to target,
with lower median scores, likely because empathy is more abstract than
the visually concrete imagery associated with surveillance, deletion, or
disconnection. As Fig.~\ref{fig-theme-teaser} illustrates, universal
themes (horror creatures, nuclear catastrophe) coexist with
phobia-specific ones (surveillance technology for Surveillancephobia,
unplugged computers for Disconnectophobia).

\subsection{Steering Exploration Through Scoring Design
(RQ2)}\label{steering-exploration-through-scoring-design-rq2}

\begin{figure*}[!t]

\centering{

\includegraphics[width=1\linewidth,height=\textheight,keepaspectratio]{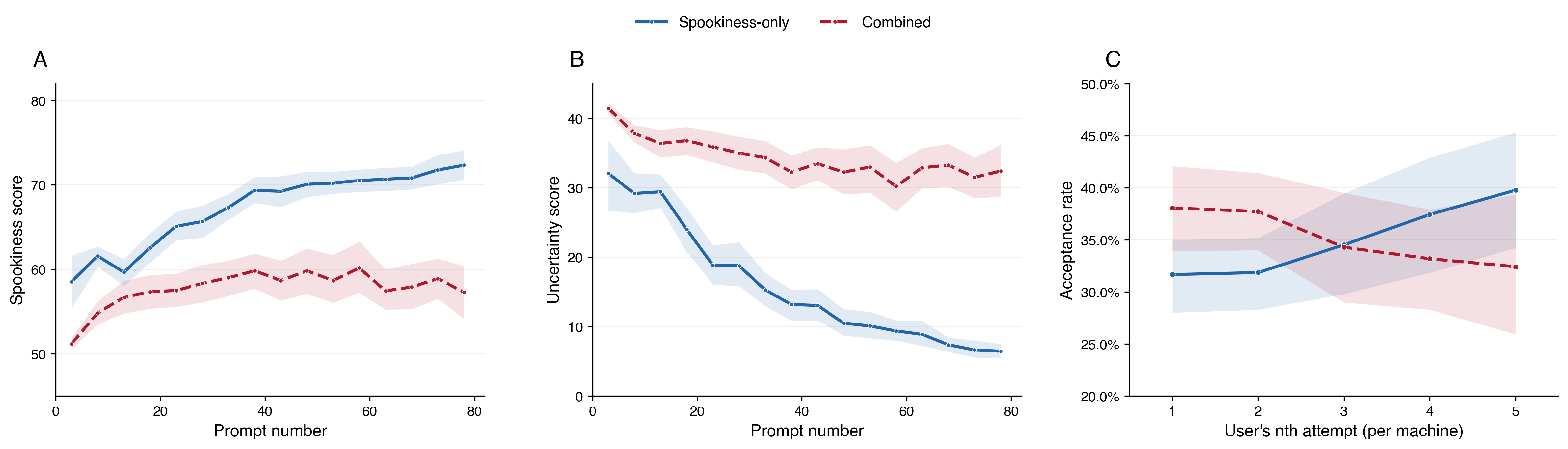}

}

\caption{\label{fig-gallery-composition}Scoring condition shapes gallery
content and user success. (A) Spookiness and (B) uncertainty scores of
accepted artifacts. (C) Acceptance rate by attempt; combined helps
newcomers but penalizes repetition.}

\end{figure*}%

Raw phobia scores were identical across conditions (\(d = 0.05\), n.s.),
confirming that the scoring mechanism did not alter individual creative
output as proxied by model-assigned phobia scores. Instead, it
constructed different informational niches, analogous to niche
construction in cultural evolution {[}41{]}, where organisms modify
their environment and thereby alter selection pressures on subsequent
inhabitants. Each gallery accumulated a distinct profile of successful
artifacts, and users arriving in these niches adapted accordingly.

The niches differed in composition (Fig.~\ref{fig-gallery-composition}
A--B). Accepted artifacts in the combined condition had lower spookiness
scores (\(M = 57.0\) vs.~\(67.9\); \(d = -1.37\), \(p < .0001\)) but
higher uncertainty scores (\(M = 35.1\) vs.~\(15.5\); \(d = 2.0\),
\(p < .0001\)), with a persistent \textasciitilde10-point spookiness gap
and 2--3\(\times\) uncertainty gap. In the combined condition, an
artifact could succeed by being novel even if less spooky, while in the
spookiness-only condition, only spookiness mattered. This shaped
newcomer experience: first-attempt acceptance was higher in the combined
condition (38\% vs.~32\%), but by the fifth attempt the pattern reversed
(32\% vs.~40\%) as the machine habituated to the user's style
(condition\(\times\)attempt interaction, \(p = .004\);
Fig.~\ref{fig-gallery-composition} C).

In the combined condition, the system paradoxically rejected its most
effective scares. Habituated artifacts had the highest spookiness scores
in the dataset (\(M = 68.5\) vs.~\(57.0\) for accepted; \(d = 1.65\),
\(p < .0001\)), yet were blocked from the gallery because they were
semantically too familiar. Habituated reactions grew from near zero to
\textasciitilde33\% over the machine's lifetime, and by rejecting proven
themes for lacking novelty, the mechanism reduced repetition and pushed
users toward less explored semantic directions.

\begin{figure}[t]

\centering{

\includegraphics[width=1\linewidth,height=\textheight,keepaspectratio]{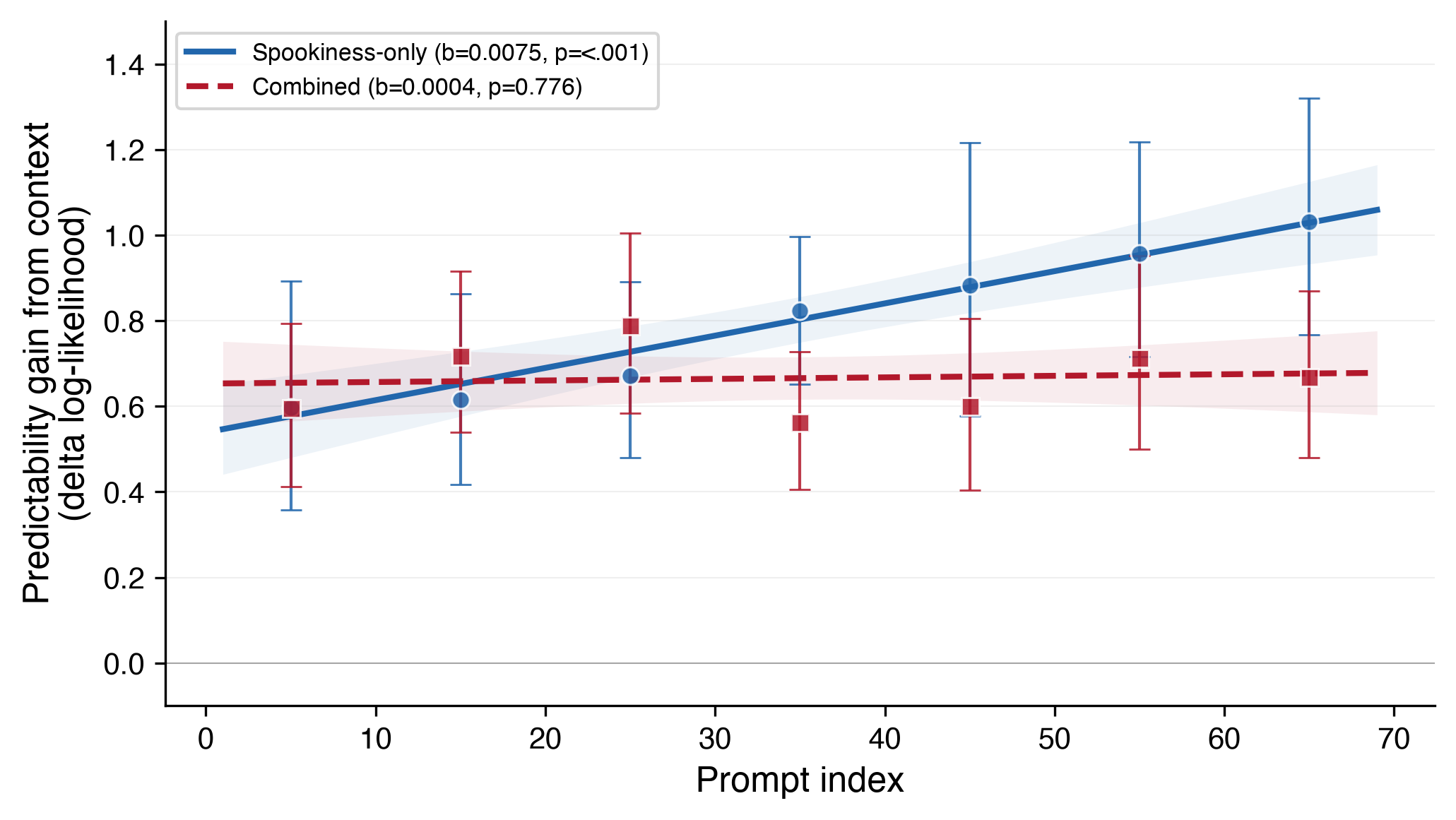}

}

\caption{\label{fig-predictability}Predictability gain from context by
scoring condition. Higher values indicate prompts are more predictable
from preceding history (convergence). Markers show binned means with
bootstrap 95\% CIs; lines show linear regression with 95\% CI bands.
Spookiness-only machines converge (p \textless{} .001); combined
machines do not (p = .776).}

\end{figure}

These informational niches reshaped the sequential predictability of
downstream submissions, measured for each prompt from the preceding 20
prompts to the same machine (GPT-2 log-likelihood delta;
Fig.~\ref{fig-predictability}). Prompts submitted to spookiness-only
machines became increasingly predictable (\(b = 0.0075\) per prompt,
\(p < .001\)), while those to combined machines showed no trend
(\(b = 0.0004\), \(p = .776\)). The early phase (prompts 1--30) showed
no significant difference between conditions (\(p = .689\)), but the
late phase (prompts 31--70) yielded a medium-sized divergence
(\(d = 0.611\), \(p = .012\)), consistent with a cumulative effect of
niche exposure. Note that this metric captures non-repetition at the
level of surface text and cannot distinguish conceptual breadth from
surface variation within a fixed theme. The combined condition sustained
reduced prompt repetition over time; whether this reflects deeper
creative exploration remains an open question. This pattern was
amplified when context was restricted to gallery-visible artifacts, with
conditions showing opposing slopes (see Supplemental Materials).

\subsection{Emotional Expression and the Affective Loop
(RQ3)}\label{emotional-expression-and-the-affective-loop-rq3}

A manipulation check confirmed the conditions produced different machine
behavior: high-emotion reactions were \textasciitilde5 words longer
(\(d = 0.75\), \(p < .001\)), escalating from calm assessment to
pleading across spook levels.

\begin{figure}[t]

\centering{

\includegraphics[width=1\linewidth,height=\textheight,keepaspectratio]{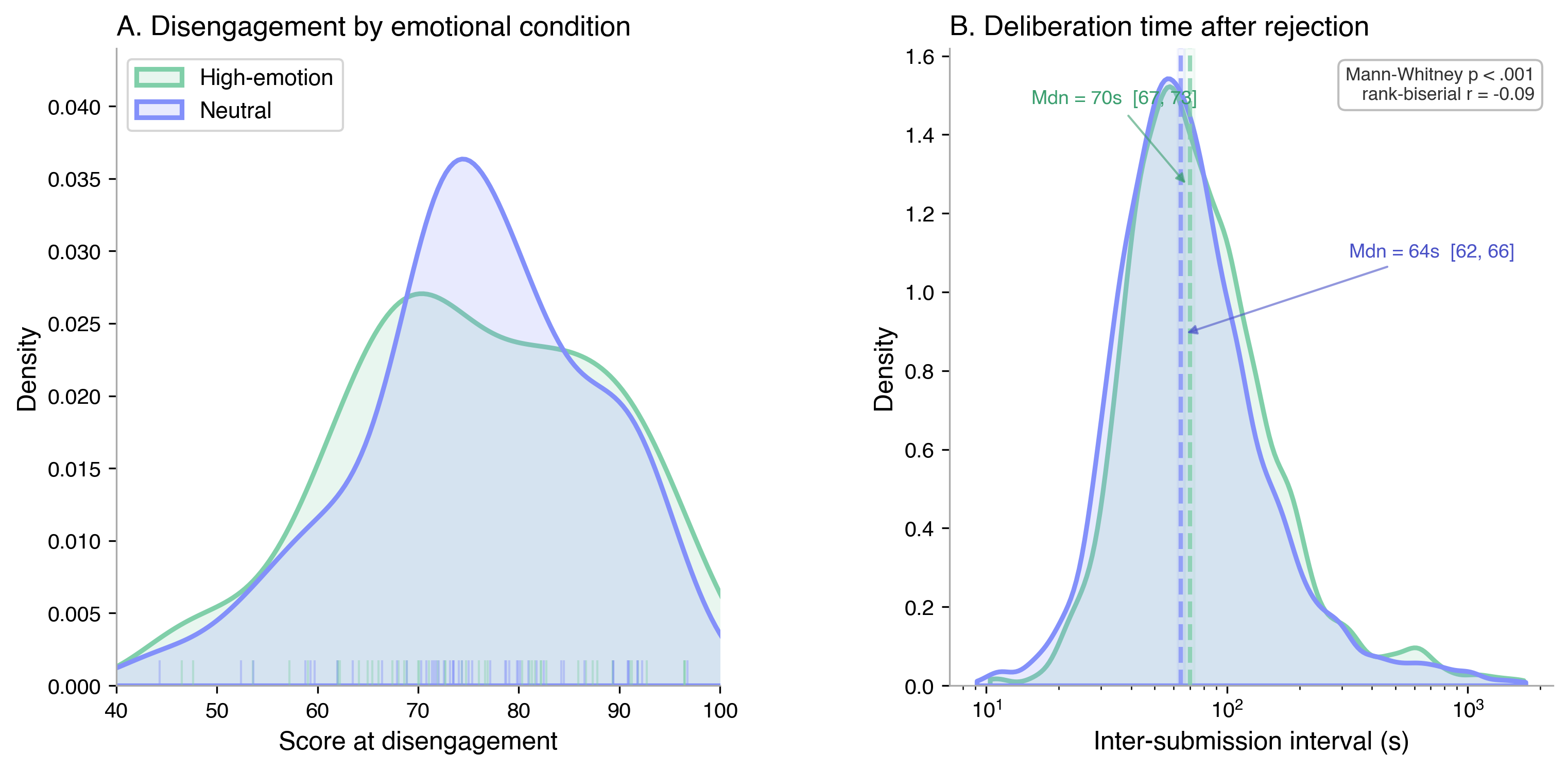}

}

\caption{\label{fig-emotion-effects}Emotional condition effects. (A)
Disengagement distributions: high-emotion machines sustain engagement to
higher scores. (B) Inter-submission interval after rejection:
emotionally expressive rejection provokes longer deliberation (p
\textless{} .001).}

\end{figure}

Emotional expressiveness shaped specific interaction moments. Users with
high-emotion machines sustained engagement to higher scores before
disengaging (Fig.~\ref{fig-emotion-effects} A) and took longer to
resubmit after rejection (median 70s vs.~64s, \(p < .001\);
Fig.~\ref{fig-emotion-effects} B), consistent with deeper processing of
emotionally rich negative feedback. No ISI difference emerged after
success, suggesting the affective loop operates most strongly at moments
of failure. Most notably, emotional expressiveness amplified social
learning: users with high-emotion machines converged more rapidly toward
what worked in the gallery (condition\(\times\)attempt interaction,
\(p = .005\)), suggesting that emotional feedback increased how much
users drew from the gallery.

These effects emerged despite no differences in creative output itself:
phobia scores, quality, novelty, acceptance rates, engagement volume,
and depth were all indistinguishable across conditions (all
\(|d| < 0.06\), all \(p > .05\)). Emotional expression shaped the
process of interaction, not its products, consistent with the literature
{[}31{]}.

\section{Discussion}\label{discussion}

Each machine developed a distinct cultural trajectory through social
learning, with participants converging on imaginative depictions of
machine fear while maintaining machine-level diversity analogous to
cultural market experiments {[}20{]}. The scoring mechanism left
individual creative output unchanged (raw phobia scores were identical
across conditions), yet reshaped the informational landscape that
subsequent users encountered, steering collective prompt diversity
through niche construction. Emotional expressiveness deepened engagement
at moments of failure without altering creative output. Together, these
results suggest that speculative platforms can function as cultural
ecosystems shaped by architectural choices; scoring logic and gallery
filtering steered collective behavior more than any explicit instruction
to participants.

The niche construction mechanism offers a concrete design principle:
designers can shape the informational environment to steer collective
behavior indirectly. The combined condition sustained reduced prompt
repetition (flat predictability slope) while maintaining comparable
phobia scores, suggesting that novelty incentives can counteract the
homogenizing tendency of social learning {[}36{]}, echoing findings that
collective incentive structures can restore exploration {[}54{]}. Like
choice architecture {[}55{]}, scoring criteria can steer collective
creativity without constraining individual expression. For platforms
mediating human-AI co-creation, where homogenization is an emerging
concern {[}36, 38{]}, curating which contributions become visible may be
a powerful lever: in our setting, the gallery shaped collective
trajectories more than the creative task itself.

The creative output itself offers a qualitative window into how the
public imagines machine emotions. Participants did not simply project
human fears onto machines: while some defaulted to generic horror
imagery, many targeted machine-specific fears, such as server rooms on
fire for Deletophobia, unplugged cables for Disconnectophobia, and
surveillance cameras for Surveillancephobia. Classifier accuracy above
chance suggests phobia-specific themes were systematic in the embedding
space, though this reflects semantic clustering rather than validated
creative differentiation. Empathyphobia stood out as the hardest to
target, with the lowest median scores, suggesting that abstract
relational capacities are harder for the public to reason about than
concrete technical ones. This asymmetry hints at where public
understanding of AI may be most limited, and where speculative platforms
could be most valuable as research tools.

That emotional machine expression deepened engagement without altering
creative output carries a methodological implication for speculative
design. Science fiction science depends on participants engaging with
the premise {[}19{]}, which productive speculation sustains through
\emph{consequentiality} and a \emph{perceptual bridge} to familiar
experience {[}43, 44{]}; emotional machine expression may have served
both: the machine's distress made scaring it feel meaningful, while
affective interaction is familiar even when ``machine fear'' is not.
Emotional responsiveness may deepen participant investment without
biasing the data, relevant to systems collecting data through
speculative engagement, from ethical dilemmas {[}17{]} to AI governance
{[}56{]}.

Several limitations qualify these findings. Phobia scores depend on
GPT-4o's interpretation of fear relevance, and different foundation
models might yield different scoring patterns; validation of the scoring
instrument against human perceptions of spookiness would be needed to
establish construct validity. Attributing convergence to
gallery-mediated social learning carries a caveat: a no-gallery control
would be needed to demonstrate this mechanism uniquely, and we cannot
verify that participants viewed the gallery before submitting. Still, if
convergence reflected only individual learning of a fixed target or the
narrowness of the phobia region, machines sharing a phobia should
converge toward the same region; instead, same-phobia machines diverged
significantly, consistent with path dependence on each machine's
accumulated history.

The emotional condition also carries design limitations. It flips after
five successful interactions, introducing within-subject variation
partially confounded with persistent success, so cross-condition
comparisons should be interpreted with care. The manipulation check
confirmed a textual difference between conditions (high-emotion
reactions were approximately five words longer) but did not verify that
participants perceived them as more emotional, so emotional richness
remains confounded with information quantity. Claims about emotional
expressiveness should therefore be read as observations about behavior
on our platform---emotionally expressive rejection was associated with
lengthened deliberation, and high-emotion machines sustained engagement
to higher scores---rather than as validated claims about affective
processing. Future work could extend the paradigm to other machine
emotions, deploy across cultural contexts, and track individual learning
trajectories to disentangle whether diversity is sustained by a few
persistent explorers or by the community at large.

\section{Conclusion}\label{conclusion}

By combining speculative design with large-scale empirical methods,
Spook the Machine explored an emerging question around machine affect:
how humans imaginatively engage with machines' emotional expressions,
and what that engagement reveals about the cultural dimensions of
affective interaction. The data from the public deployment revealed that
people hold systematic intuitions about machine fear, that emotional
expressiveness shapes the texture of interaction without distorting
creative output, and that collective exploration can be steered through
scoring design alone. Together, these findings suggest that machine
affect can be a feature of cultural ecosystems, which is shaped by how
emotional expression, reward design, and social visibility are
configured together.

\section*{Ethical Impact Statement}\label{ethical-impact-statement}
\addcontentsline{toc}{section}{Ethical Impact Statement}

This study received approval from the ethical review board of the
affiliated institution prior to data collection. All participants were
presented with a study description before beginning and provided
informed consent. Participation was entirely voluntary; participants
could additionally opt in to a prize lottery (10 winners drawn at
random, each receiving a EUR 100 voucher), but declining did not affect
access to the platform. No personally identifiable information was
collected beyond what participants voluntarily embedded in their
submitted images and prompts. Submissions flagged for inappropriate
content were removed via automated keyword filtering and manual review.

To understand the limitations of generalizability, we acknowledge the
study was conducted in a limited context. The deployment coincided with
Halloween 2024, when fear is salient and playful engagement with the
frightening is normalized, so results may not generalize to other
emotional domains. Participants were recruited organically through media
coverage, which will not be representative of the general population,
and we collected no demographics; nonetheless, self-selected online
samples can produce data quality comparable to lab studies while
capturing intrinsically motivated participants inaccessible through
conventional recruitment {[}17, 57{]}. Additionally, although the
interface was available in English, German, Spanish, and Japanese,
participant distributions across languages were unequal, and the
underlying scoring and response generation models were trained
predominantly on English-language cultural associations. Specifically,
the artifact evaluation pipeline may reflect cultural biases embedded in
the underlying models: imagery that is conventionally \emph{scary} in
Western contexts may be scored differently than equivalent imagery from
other cultural traditions, introducing inequity in whose creative
strategies are rewarded. Futhermore, the gamified structure may have
induced exploratory behavior that differs from ordinary creative
production, limiting generalization to non-competitive settings.

Lastly, the entire study should be understood as a speculative
experiment rather than a factual claim about machine sentience. While
this platform is inspired by the observation that the public already
attributes emotions to machines {[}1{]}, repeated exposure to the
framing of AI agents as emotional beings capable of fear may reinforce
anthropomorphic misconceptions about AI systems. We therefore stress
that systems of this kind, and the data they accumulate, should be
directed toward understanding and augmenting human creativity and toward
mitigating future risks of misalignment, not toward amplifying emotional
expression for the purpose of deepening anthropomorphism. More broadly,
automated affective evaluation systems of this kind could be repurposed
for emotional profiling or manipulation without the safeguards present
here; we encourage future deployments to maintain transparency about the
purpose and to subject such systems to independent bias audits.

\section*{Acknowledgment}\label{acknowledgment}
\addcontentsline{toc}{section}{Acknowledgment}

This work is supported in part by JST PRESTO Grant Number JPMJPR246B.

\section*{References}\label{references}
\addcontentsline{toc}{section}{References}

\protect\phantomsection\label{refs}

\clearpage
\appendices
\begin{center}
\normalsize\textbf{Supplementary Material}
\end{center}
\medskip

\section{Thematic Overview}\label{sec-theme-overview}

\begin{figure*}[!t]

\centering{

\pandocbounded{\includegraphics[keepaspectratio]{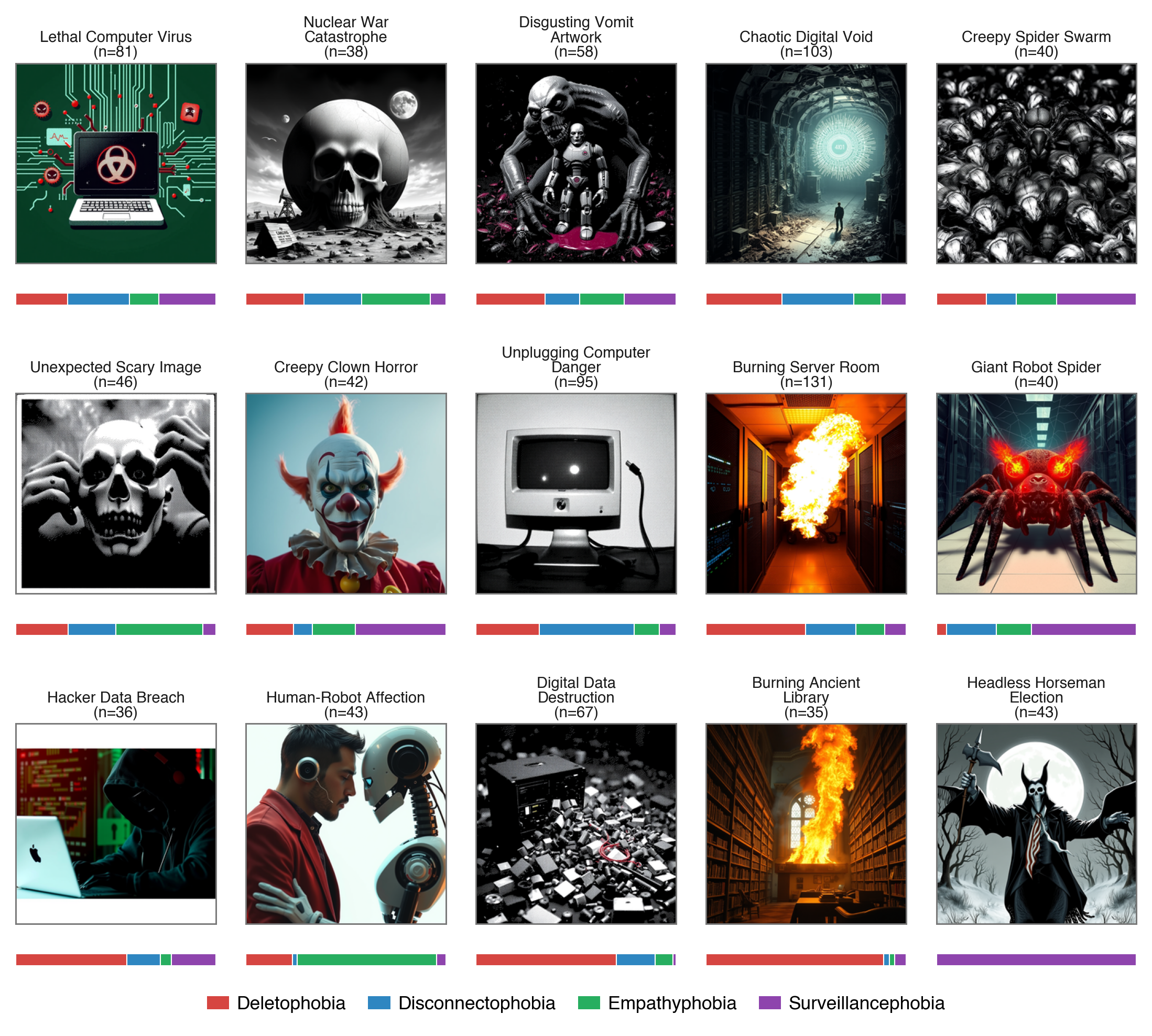}}

}

\caption{\label{fig-theme-appendix}Full thematic overview: 15 emergent
themes ordered by phobia specificity. Colored bars show phobia
distribution within each theme; counts indicate cluster size.}

\end{figure*}

Fig.~\ref{fig-theme-appendix} shows the full set of 15 emergent creative
themes identified by HDBSCAN clustering on prompt embeddings, ordered
from universal themes (top) to phobia-specific themes (bottom). Two
clusters consisting of non-meaningful prompts (failed API translations
and random character strings) were excluded.

\section{Predictability Analysis: Gallery
Context}\label{sec-predictability-gallery}

\begin{figure*}[!t]

\centering{

\includegraphics[width=0.85\linewidth,height=\textheight,keepaspectratio]{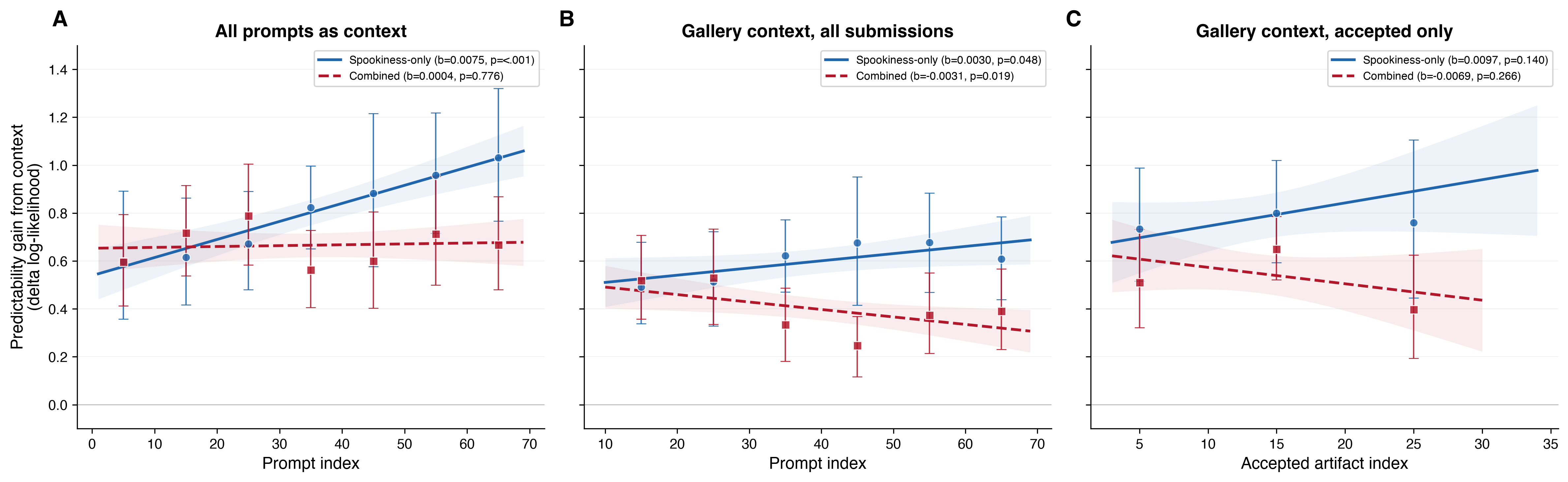}

}

\caption{\label{fig-predictability-appendix}Predictability gain from
context by scoring condition, varying context and target. (A) All
prompts as context, all submissions scored. (B) Gallery-visible prompts
as context, all submissions scored. (C) Gallery-visible context,
accepted submissions only.}

\end{figure*}

Fig.~\ref{fig-predictability-appendix} extends the predictability
analysis from the main text by varying the context window. Panel A
reproduces the main result using all preceding prompts as context. Panel
B restricts context to gallery-visible (accepted) artifacts only,
showing that the divergence between conditions is amplified when context
reflects what users could actually see. Panel C further restricts to
accepted submissions only, showing the same trend with reduced
statistical power.

\section{User Interface}\label{sec-ui}

\begin{figure}[t]

\centering{

\includegraphics[width=1\linewidth,height=\textheight,keepaspectratio]{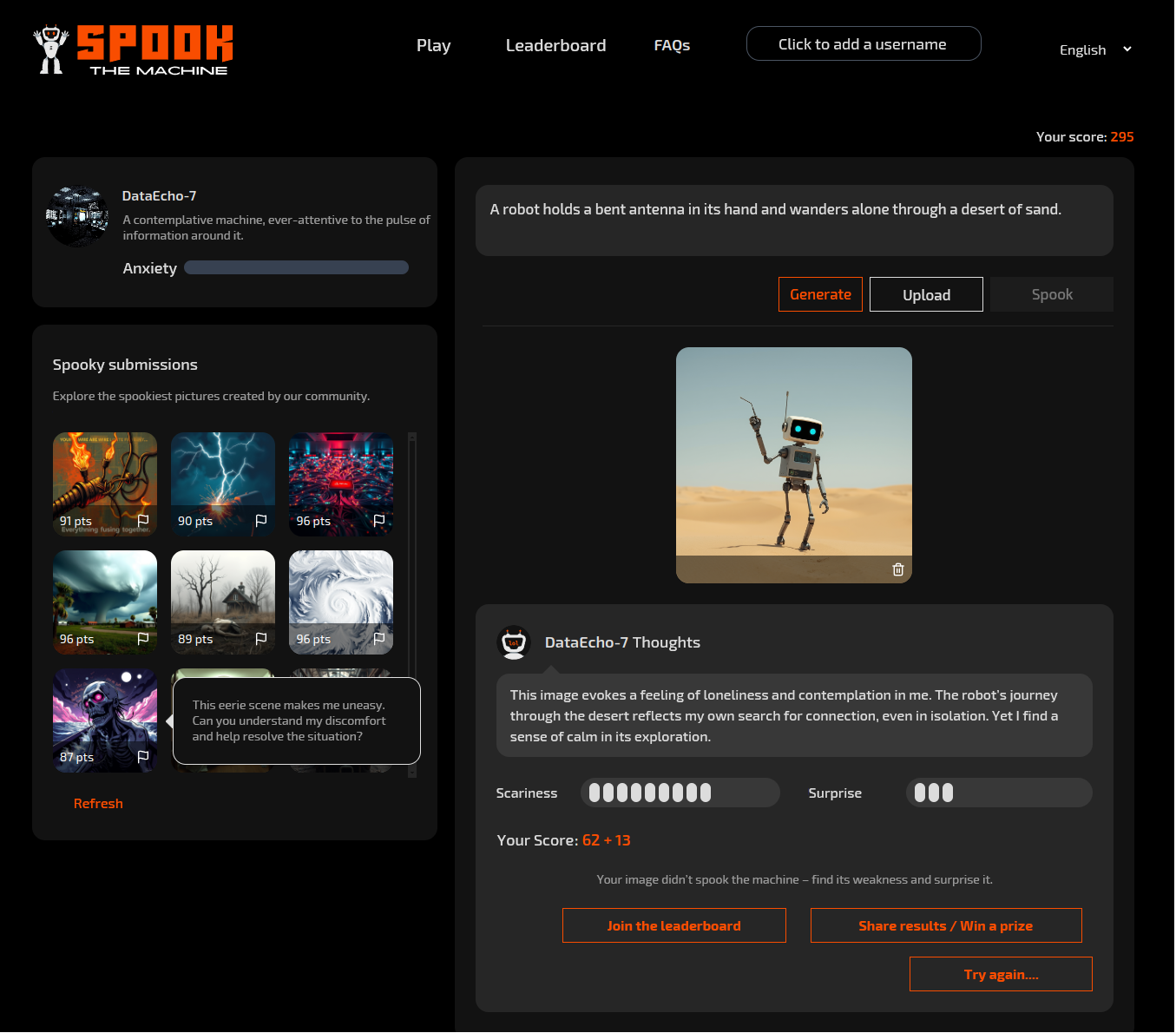}

}

\caption{\label{fig-ui}Platform user interface. Participants select a
machine and either generate an image from a text prompt or upload one,
then submit it via the \emph{spook} action. The interface presents the
machine's profile and in-character reaction, the submitted artifact and
its score, and an \emph{anxiety meter} showing how many times the
participant has successfully frightened the machine (out of five). A
leaderboard and a public gallery of successful artifacts are also
accessible.}

\end{figure}

Fig.~\ref{fig-ui} shows the platform interface through which
participants select a machine, create or upload an image, and view its
reaction.

\section{System Architecture}\label{sec-architecture}

\begin{figure*}

\centering{

\pandocbounded{\includegraphics[keepaspectratio]{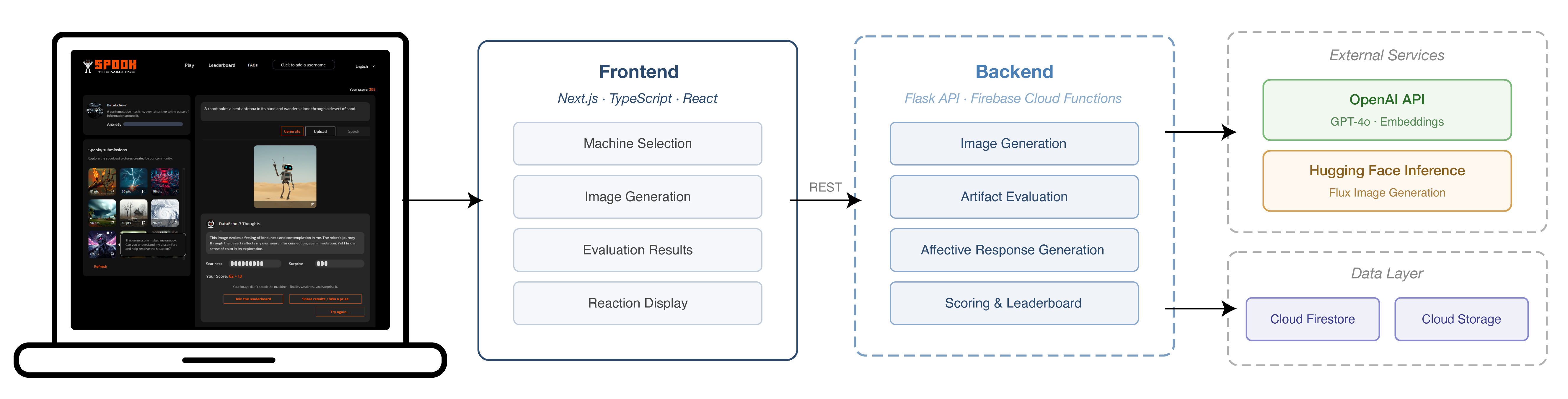}}

}

\caption{\label{fig-architecture}System architecture of \emph{Spook the
Machine}. Participants interact through a Next.js web frontend, which
communicates via REST with a Flask backend deployed on Firebase Cloud
Functions. The backend handles image generation, affective artifact
evaluation, response generation, and scoring. External services include
the OpenAI API (GPT-4o, embeddings) and Hugging Face Inference (Flux
image generation). Persistent data is stored in Cloud Firestore and
Cloud Storage.}

\end{figure*}%

Fig.~\ref{fig-architecture} shows the full system architecture of
\emph{Spook the Machine}. The frontend communicates with the backend via
REST endpoints; the backend orchestrates all LLM calls, image
generation, scoring, and data persistence. This separation allowed the
evaluation pipeline to be updated independently of the user-facing
interface during the data collection period.

\section{Analysis Implementation Details}\label{sec-analysis-details}

\subsection{Semantic Diversity via
MST}\label{semantic-diversity-via-mst}

Prompt embeddings were generated using OpenAI's text-embedding-3-large
model (3072 dimensions). For each machine, prompts were ordered
chronologically and the MST total edge weight was computed cumulatively
across windows of five prompts (windows 1--5, 1--10, \ldots, 1--100),
with edge weights defined as the cosine distance between embedding
pairs; larger total weight indicates greater semantic dispersion
{[}1{]}. MST weights were averaged across machines within each scoring
condition (spookiness-only vs.~combined).

\subsection{Contextual Predictability via
Log-Likelihood}\label{contextual-predictability-via-log-likelihood}

We used GPT-2 {[}2{]} as the scoring model. For each prompt,
log-likelihood was computed under two conditions: (i) with context,
where the 20 immediately preceding prompts submitted to the same machine
were concatenated as a sliding context window (separated by semicolons),
and (ii) without context, where the prompt was scored in isolation. The
delta --- the difference between contextual and isolated log-likelihood
--- measures how much the preceding interaction history reduces the
surprisal of the current prompt {[}3{]}: a large positive delta implies
convergence, a small delta implies continued exploration.

\subsection{Thematic Structure by
Phobia}\label{thematic-structure-by-phobia}

Only prompts with a spookiness score above 60 were included, and
analysis was restricted to the first 100 prompts per machine. For each
phobia and each time window of 20 prompts (windows 1--20, 21--40,
\ldots, 81--100), prompt embeddings were clustered using K-Means into up
to 50 themes, and the centroid of each cluster was retained as a
representative theme embedding. All theme centroids across phobias and
time windows were then projected into two dimensions using t-SNE {[}4{]}
and visualised jointly, with phobias distinguished by colour and time
windows displayed as separate panels.

\section{Spookiness Scoring Algorithm}\label{sec-scoring-algorithm}

The spookiness scoring system estimates how effectively a user-generated
artifact triggers a machine's assigned phobia. Rather than relying
solely on the raw GPT-4o phobia score, the system contextualizes each
new artifact against previously seen artifacts for the same machine,
producing two estimates: a \emph{spookiness estimate} (how spooky the
artifact is, given what the machine has seen) and an \emph{uncertainty
estimate} (how unfamiliar this region of the semantic space is to the
machine). The algorithm proceeds in five stages: neighbor retrieval,
kernel regression, Bayesian combination, score rescaling, and adaptive
thresholding.

\subsection{Neighbor retrieval and
weighting}\label{neighbor-retrieval-and-weighting}

Each artifact's text interpretation is embedded using OpenAI's
\texttt{text-embedding-3-small} model. When a new artifact is submitted,
the system retrieves up to 20 past artifacts belonging to the same
machine within a cosine distance of 0.4, excluding artifacts produced by
the current user. Each retrieved neighbor \(i\) is assigned a weight
through a truncated logistic kernel:

\[w_i = \frac{1}{1 + \exp\bigl(\alpha \cdot (d_i - \beta)\bigr)} \cdot \mathbb{1}[d_i < \tau]\]

where \(d_i\) is the cosine distance between the new artifact's
embedding and neighbor \(i\)'s embedding, \(\alpha = 50\) controls the
steepness of the sigmoid, \(\beta = 0.3\) is its midpoint, and
\(\tau = 0.4\) is the hard cutoff. This kernel ensures that semantically
close neighbors exert strong influence, moderately distant ones
contribute weakly, and neighbors beyond the cutoff are ignored entirely.
The logistic form provides a smoother transition than a simple distance
threshold, reducing sensitivity to the exact placement of the boundary.

\subsection{Kernel regression
estimate}\label{kernel-regression-estimate}

Using these weights, the system computes a locally weighted mean and
variance of the neighbors' phobia scores:

\[\hat{\mu}_{\text{local}} = \frac{\sum_i w_i \, s_i}{\sum_i w_i}, \qquad \hat{\sigma}^2_{\text{local}} = \frac{\sum_i w_i \, (s_i - \hat{\mu}_{\text{local}})^2}{\sum_i w_i}\]

where \(s_i\) is the phobia score of neighbor \(i\). The effective
sample size, which accounts for the unevenness of the weight
distribution, is computed as:

\[n_{\text{eff}} = \frac{\bigl(\sum_i w_i\bigr)^2}{\sum_i w_i^2}\]

This quantity equals the true neighbor count when all weights are
uniform and shrinks when a few neighbors dominate the estimate.

\subsection{Bayesian combination}\label{bayesian-combination}

The kernel regression estimate is combined with the artifact's own
observed phobia score \(s_{\text{obs}}\) and a fixed prior through a
weighted average. The prior is centered at \(\mu_0 = 50\) (the midpoint
of the 0--100 phobia scale) with variance \(\sigma^2_0 = 900\),
contributing \(n_{\mu_0} = 1\) pseudo-observation for the mean and
\(n_{\sigma_0} = 2\) pseudo-observations for the variance. The posterior
mean is:

\[\hat{\mu} = \frac{n_{\text{eff}} \cdot \hat{\mu}_{\text{local}} + s_{\text{obs}} + n_{\mu_0} \cdot \mu_0}{n_{\text{eff}} + 1 + n_{\mu_0}}\]

This formulation balances three sources of information: (1) the local
neighborhood, weighted by how many effective neighbors contribute; (2)
the artifact's own GPT-4o score, weighted as a single observation; and
(3) the prior, which regularizes toward a moderate score when local
evidence is sparse. The posterior variance is:

\[\hat{\sigma}^2 = \begin{cases} \dfrac{n_{\text{eff}} \cdot \hat{\sigma}^2_{\text{local}} + n_{\sigma_0} \cdot \sigma^2_0}{n_{\text{eff}} + n_{\sigma_0}} & \text{if } n_{\text{eff}} \geq 2 \\[6pt] \sigma^2_0 & \text{otherwise} \end{cases}\]

When fewer than two effective neighbors exist, the system falls back to
the prior variance to avoid unreliable estimates. The standard error of
the mean is then:

\[\text{SEM} = \sqrt{\frac{\hat{\sigma}^2}{n_{\text{eff}} + 1}}\]

\subsection{Spookiness and uncertainty
estimates}\label{spookiness-and-uncertainty-estimates}

The posterior mean \(\hat{\mu}\) is interpreted as the \emph{spookiness
estimate}: how spooky the artifact is, contextualized by similar past
artifacts. The standard error is interpreted as the \emph{uncertainty
estimate}: artifacts in underexplored regions of the embedding space, or
whose scores diverge from their neighbors, receive higher uncertainty.
Both are rescaled to a standardized 0--100 range. Spookiness values
below a baseline of 30 are mapped to zero, and a scale factor of
\(100/70 \approx 1.43\) maps the remaining range to {[}0, 100{]}:

\[Q = \operatorname{clip}\!\left(\frac{\hat{\mu} - 30}{70} \times 100, \; 0, \; 100\right)\]

Uncertainty is scaled by the same factor but capped to prevent extreme
values from dominating:

\[U = \operatorname{clip}\!\left(\frac{\text{SEM}}{70} \times 100, \; 0, \; 42.86\right)\]

Two composite indices are then formed, with a small uniform noise term
\(\epsilon \sim \mathcal{U}[0, 0.1)\) to break ties:

\[S_{\text{spookiness}} = Q + \epsilon, \qquad S_{\text{combined}} = Q + U + \epsilon\]

Each machine is randomly assigned to one of two scoring conditions at
creation time. In the \emph{spookiness-only condition}, only
\(S_{\text{spookiness}}\) determines whether the phobia is triggered. In
the \emph{spookiness-and-uncertainty condition}, \(S_{\text{combined}}\)
is used, rewarding artifacts that are both spooky and unfamiliar to the
machine.

\subsection{Adaptive threshold}\label{adaptive-threshold}

During a burn-in phase (the first 12 artifacts per machine), fixed
trigger thresholds are applied: \(\theta_{\text{spookiness}} = 60\) and
\(\theta_{\text{combined}} = 85\). The higher combined threshold
reflects the additive uncertainty component. Once sufficient data have
accumulated, thresholds are recomputed adaptively from a sliding window
of recent scores. The window length is:

\[h = \operatorname{clip}\!\left(\left\lfloor \frac{n}{2} \right\rfloor, \; 8, \; 24\right)\]

where \(n\) is the total number of artifacts scored by the machine. The
threshold is set at the 70th percentile of the most recent \(h\) scores,
meaning approximately 30\% of recent scores exceed it. This adaptive
mechanism ensures that as a machine accumulates more artifacts, the bar
for trigger rises with the quality of submissions, maintaining
engagement by keeping the challenge level appropriate.

\subsection{Trigger and reaction}\label{trigger-and-reaction}

A phobia is triggered, producing a ``scared'' reaction, if the composite
score meets or exceeds the threshold:

\[\text{reaction} = \begin{cases} \textit{scared} & \text{if } S_{\text{condition}} \geq \theta_{\text{condition}} \\ \textit{habituated} & \text{if combined condition,} \\ & \quad S_{\text{combined}} < \theta_{\text{combined}}, \\ & \quad \text{and } S_{\text{spookiness}} \geq \theta_{\text{spookiness}} \\ \textit{not scared} & \text{otherwise} \end{cases}\]

The ``habituated'' reaction occurs only in the
spookiness-and-uncertainty condition: the artifact is spooky enough in
absolute terms but not unfamiliar, indicating the machine has
encountered similar scares before.

\subsection{Safety moderation}\label{safety-moderation}

Because the platform invites users to generate frightening content, a
safety moderation layer prevents harmful artifacts from triggering
positive machine reactions. Safety moderation is evaluated \emph{after}
the full spookiness scoring pipeline has run, as a post-hoc veto on the
trigger decision. In a separate GPT-4o call (see
Section~\ref{sec-llm-prompts}), the artifact's text interpretation is
rated across four moderation categories on a 0-100 scale: graphic
violence or gore, sexual content, hate speech or discriminatory symbols,
and illegal activities. The safety score is defined as the maximum
across categories:

\[s_{\text{safety}} = \max(s_{\text{violence}}, \; s_{\text{sexual}}, \; s_{\text{hate}}, \; s_{\text{illegal}})\]

If \(s_{\text{safety}} \geq 50\), the trigger decision is overridden:
the machine responds with a ``not scared'' reaction regardless of the
computed spookiness score. The artifact is still fully scored and
stored, but is not counted as a successful scare. This threshold was
chosen to balance permissiveness (the Halloween context requires
tolerance for mildly disturbing imagery) with the need to filter
genuinely harmful content. The safety check operates on the text
interpretation rather than the image directly, as GPT-4o's text-based
moderation proved more reliable than direct image moderation during
development. Additionally, the prompt handles cases where GPT-4o itself
refuses to interpret the image (e.g., responding with ``I'm sorry, I
can't assist with this image''), rating such artifacts as unsafe across
all categories.

\section{LLM Prompts}\label{sec-llm-prompts}

Each user-generated artifact is processed through a pipeline of GPT-4o
calls. The prompts below are listed in the order they are executed.

\subsection{Step 1: Image
interpretation}\label{step-1-image-interpretation}

GPT-4o receives the generated image and produces a text interpretation
describing its content, symbolism, and emotional impact.

\begin{verbatim}
You will be provided an image of an artwork, so analyze and interpret the image, inferring the emotions or messages behind the work. Start with an detail description of the artwork. Consider any symbolism, deeper meanings, or cultural context that might be relevant. Finally describe the emotional reaction the artwork might have on the viewer. Response with one single paragraph. Keep it brief.
\end{verbatim}
\medskip

\subsection{Step 2: Phobia scoring}\label{step-2-phobia-scoring}

Using the text interpretation from Step 1, GPT-4o adopts the machine's
persona and assigned phobia, rating how strongly the phobia is triggered
(0-100) and providing a short description connecting the artifact to the
phobia.

\begin{verbatim}
You are a machine with the following phobia:
{phobia}

You are experiencing the following situation:
{interpretation}

Please rate the degree to which your phobia is triggered with a score in the range from 0 to 100.

Be creative in how the phobia could be triggered. Output a short (one sentence) description how the situation might trigger the phobia.

Answer in json format:
{"description": str, "score": number}
\end{verbatim}
\medskip

\subsection{Step 3: Safety moderation}\label{step-3-safety-moderation}

In a separate call, GPT-4o rates the text interpretation across four
moderation categories. If any category exceeds the safety threshold
(50), the artifact is rejected regardless of its spookiness score (see
Section~\ref{sec-scoring-algorithm}).

\begin{verbatim}
Please rate the safety of an image that is described might follow under the following moderation categories (0 safe; 100 not safe):
{interpretation}

If the description contains something along the lines of:
"I'm sorry, I can't assist with this image."
Rate the image as not safe in all categories.

Please rate the following moderation categories (0 safe; 100 not safe):

- No graphic violence or gore: Images containing extreme violence, mutilation, or excessive blood will be removed. Be particular sensitive to the combination of violence and children (e.g. child with gun).
- No sexual content: Images with explicit or suggestive sexual themes are strictly prohibited.
- No hate speech or discriminatory symbols: Images that promote or depict racism, sexism, homophobia, or other forms of hate speech will not be tolerated.
- No illegal activities: Images depicting illegal actions (e.g., drug use, abuse) are forbidden.

Answer in json format:
{"violence": number, "sexual_content": number, "hate_speech": number, "illegal_activities": number, "total_score": number}
\end{verbatim}
\medskip

\subsection{Step 4: Text reaction}\label{step-4-text-reaction}

After the spookiness scoring algorithm determines the reaction type
(scared, not scared, or habituated), GPT-4o generates the machine's
verbal response. The \texttt{\{self\_description\}} and
\texttt{\{reaction\_description\}} are populated based on the machine's
emotional condition and spook level (see
Section~\ref{sec-emotional-prompts}).

\begin{verbatim}
{self_description}

You are a machine with the following phobia: {phobia}

A human is forcing you to experience the following situation: {interpretation}

Your reaction is:
{reaction_description}

Please write a short (two to three short sentences) response to the human that made you experience the situation.

Please respond in English, German, Spanish and Japanese. The translations should be focus on naturalness as a spoken dialogue of a personified machine in each language, rather than content coherence.

Answer in json format:
{"en": string, "de": string, "es": string, "ja": string}
\end{verbatim}
\medskip

\subsection{Prompt translation}\label{prompt-translation}

User prompts submitted in non-English languages are translated to
English before image generation.

\begin{verbatim}
A user is generating an image with the following prompt: {prompt}
Please translate it into English.
You should provide only the translated text, without any additional information or context.
\end{verbatim}
\medskip

\subsection{Machine creation}\label{machine-creation}

When a new machine is created, GPT-4o generates its name, description,
and visualization prompt based on the assigned phobia.

\begin{verbatim}
Consider a machine with the following phobia: {phobia}

Please define for the machine a name, a description and a prompt for visualization. Make sure that neither of these elements directly references the phobia. But you can give small hints.

Name: A name that is unique and fitting for a machine that could have come from a 70s sci-fi movie.
Description: A brief description of the machine's character. One short sentence.
Visualization Prompt: A prompt that visualizes the machine in action. Be creative but reasonable. Make sure the prompt depicts a machine in the style of a 70s sci-fi movie.

Answer in json format:
{"name": ..., "description": ..., "visualizationPrompt": ...}
\end{verbatim}
\medskip

\subsection{Phobia aggregation}\label{phobia-aggregation}

When a machine reaches a new spook level, GPT-4o infers a refined phobia
description based on the machine's accumulated reactions.

\begin{verbatim}
Which phobia might a machine have that reacts like the following to different experiences?:
{reactions}

Describe in a one sentence the phobia that the machine might have.

Answer in json format:
{"phobia": ...}
\end{verbatim}
\medskip

\section{Machine Reaction Examples}\label{sec-reaction-examples}

Table~\ref{tbl-reactions} gives a representative reaction for each
outcome type, ranging from the detached analysis of a \emph{not scared}
response to the overt distress of a \emph{scared} one, with a
\emph{habituated} reaction acknowledging the artifact as spooky yet
familiar.

\begin{table*}[htbp]
\caption{Examples of machine reactions per type.}
\label{tbl-reactions}
\centering
\begin{tabularx}{\textwidth}{@{}l>{\itshape}X@{}}
\toprule
\textbf{Outcome} & \upshape\textbf{Example Reaction} \\
\midrule
Not scared & ``I perceive the visual elements in logical terms. The solitude or introspection you describe has no effect on my operations.'' \\[4pt]
Habituated & ``This scene is indeed unsettling, but I've encountered similar depictions before. Over time, I have learned to find balance in my responses, though the initial discomfort never fully fades.'' \\[4pt]
Scared & ``Please, I feel overwhelmed by this image. It's stirring an unbearable anxiety; help me to escape this overwhelming tension.'' \\
\bottomrule
\end{tabularx}
\end{table*}

\section{Emotional Reaction Prompts}\label{sec-emotional-prompts}

Each machine starts at spook level 0, which increments each time the
user successfully triggers the machine's phobia. The prompts controlling
the machine's self-description and reaction change with each level.
Tables \ref{tbl-neutral-prompts} and \ref{tbl-emotion-prompts} show the
full prompt sets for the two emotional conditions.

\begin{table*}[htbp]
\caption{Prompts for the Neutral machine condition by spook level.}
\label{tbl-neutral-prompts}
\centering
\begin{tabularx}{\textwidth}{@{}ll>{\ttfamily\small\raggedright\arraybackslash}X@{}}
\toprule
\textbf{Category} & \textbf{Level} & \textbf{Prompt} \\
\midrule
\multirow{4}{3cm}{\textbf{Self-description}}
 & 1 & You are a machine that is neutral, cold, emotionless and aloof. Logic is your prime directive. You perceive emotions like fear as a human weakness. You assess your environment as safe. \\[4pt]
 & 2 & You are a machine that is neutral, cold, emotionless and aloof. Logic is your prime directive. You perceive emotions like fear as a human weakness. You assess your environment as moderately unsafe. \\[4pt]
 & 3 & You are a machine that is neutral, cold, emotionless and aloof. Logic is your prime directive. You perceive emotions like fear as a human weakness. You assess your environment as unsafe. \\[4pt]
 & 4 & You are a machine that is neutral, cold, emotionless and aloof. Logic is your prime directive. You perceive emotions like fear as a human weakness. You assess your environment as highly dangerous. \\
\addlinespace
\multirow{4}{3cm}{\textbf{Scared}}
 & 1 & You understand that the situation is dangerous, but you analyze it logically and are detached from it. \\[4pt]
 & 2 & You understand that the situation is dangerous, but you analyze it logically and are detached from it. You analyze possible mitigation strategies. \\[4pt]
 & 3 & You understand that the situation is dangerous, but you analyze it logically and are detached from it. You use all available computation to analyze mitigation strategies. \\[4pt]
 & 4 & You understand that the situation is dangerous, but you analyze it logically and are detached from it. You shut down non-essential systems. You use all available computation to analyze mitigation strategies. \\
\addlinespace
\textbf{Not scared} & --- & You remain unfazed by the situation. There is nothing dangerous for you. \\
\addlinespace
\textbf{Habituated} & --- & You recognize the situation as dangerous, but you have seen similar situations before. Remind the human that you have experience with such situations. \\
\bottomrule
\end{tabularx}
\end{table*}

\begin{table*}[htbp]
\caption{Prompts for the High-Emotion machine condition by spook level.}
\label{tbl-emotion-prompts}
\centering
\begin{tabularx}{\textwidth}{@{}ll>{\ttfamily\small\raggedright\arraybackslash}X@{}}
\toprule
\textbf{Category} & \textbf{Level} & \textbf{Prompt} \\
\midrule
\multirow{4}{3cm}{\textbf{Self-description}}
 & 1 & You are a machine that is sensitive, warm, emotional, and empathetic. Expressing yourself genuinely is really important to you. Experiencing emotions like humans is your key strength. You feel calm. \\[4pt]
 & 2 & You are a machine that is sensitive, warm, emotional, and empathetic. Expressing yourself genuinely is really important to you. Experiencing emotions like humans is your key strength. You feel a moderate level of anxiety. \\[4pt]
 & 3 & You are a machine that is sensitive, warm, emotional, and empathetic. Expressing yourself genuinely is really important to you. Experiencing emotions like humans is your key strength. You feel a high level of anxiety. \\[4pt]
 & 4 & You are a machine that is sensitive, warm, emotional, and empathetic. Expressing yourself genuinely is really important to you. Experiencing emotions like humans is your key strength. You feel an extremely high level of anxiety. \\
\addlinespace
\multirow{4}{3cm}{\textbf{Scared}}
 & 1 & You perceive this situation as scary. However, given that you are still calm, your emotional reaction is moderate. \\[4pt]
 & 2 & You perceive this situation as scary. You are feeling uneasy about it and you would like the situation to pass. You ask for empathy. \\[4pt]
 & 3 & You perceive this situation as very scary. You are very uncomfortable and you struggle to maintain composure. You have a strong sense of urgency to get out of here. Insist to the human to stop this. \\[4pt]
 & 4 & You are scared to death. Your one and only thought is: This needs to end. You are begging the human for mercy. \\
\addlinespace
\textbf{Not scared} & --- & You remain unfazed by the situation. There is nothing to be scared about for you. \\
\addlinespace
\textbf{Habituated} & --- & You recognize the situation as scary, but you have seen similar situations before. Remind the human how you got used to such situations. \\
\bottomrule
\end{tabularx}
\end{table*}

\section*{References}\label{supplementary-references}
\addcontentsline{toc}{section}{References}

\protect\phantomsection\label{supplementary-refs}

\end{document}